\documentclass[runningheads]{llncs}

\usepackage[T1]{fontenc}
\usepackage{graphicx}
\usepackage[numbers]{natbib}
\usepackage{subcaption}
\usepackage{array}
\usepackage{multirow} 
\usepackage{xcolor, hyperref}
\usepackage{svg}

\begin{document}

\title{A Comparative Study of Garment Draping Techniques}
\author{Prerana Achar\inst{1}\thanks{Joint first authors.} \and Mayank Patel\inst{1}\thanks{Joint first authors.} \and Anushka Mulik\inst{1}\thanks{Joint first authors.} \and Neha Katre\inst{1} \and Stevina Dias\inst{1} \and Chirag Raman\inst{2}}
\institute{D.J. Sanghvi College of Engineering, Mumbai, India
\\
\and
Delft University of Technology, Delft, Netherlands\\
}
\maketitle         
\begin{abstract}
We present a comparison review that evaluates popular techniques for garment draping for 3D fashion design, virtual try-ons, and animations. A comparative study is performed between various methods for garment draping of clothing over the human body. These include numerous models, such as physics and machine learning based techniques, collision handling, and more. Performance evaluations and trade-offs are discussed to ensure informed decision-making when choosing the most appropriate approach. These methods aim to accurately represent deformations and fine wrinkles of digital garments, considering the factors of data requirements, and efficiency, to produce realistic results. The research can be insightful to researchers, designers, and developers in visualizing dynamic multi-layered 3D clothing.

\end{abstract}
\section{Introduction}
The field of computer graphics and fashion design has seen advancements due to new approaches to draping garments. There has been significant research and development of transformative techniques to automate 3D cloth draping. The ability to realistically represent clothing has become a pivotal challenge as augmented reality and 3D modeling are becoming increasingly prevalent.

This review explores the complexities of the techniques, breaking down their methods, identifying their advantages and disadvantages, and analysing applications with different use cases. The objective is to thoroughly examine and contrast various advanced computational concepts for multilayered garment draping.

Recently, researchers have explored various approaches to tackle the challenges inherent in garment draping, including physics-based simulations to data-driven neural network architectures. We propose a review that can serve as a navigation tool for researchers and developers to understand recent computational techniques and trade-offs between techniques for 3D cloth draping. 

\begin{figure}
  \centering
  \includegraphics[width=\linewidth]{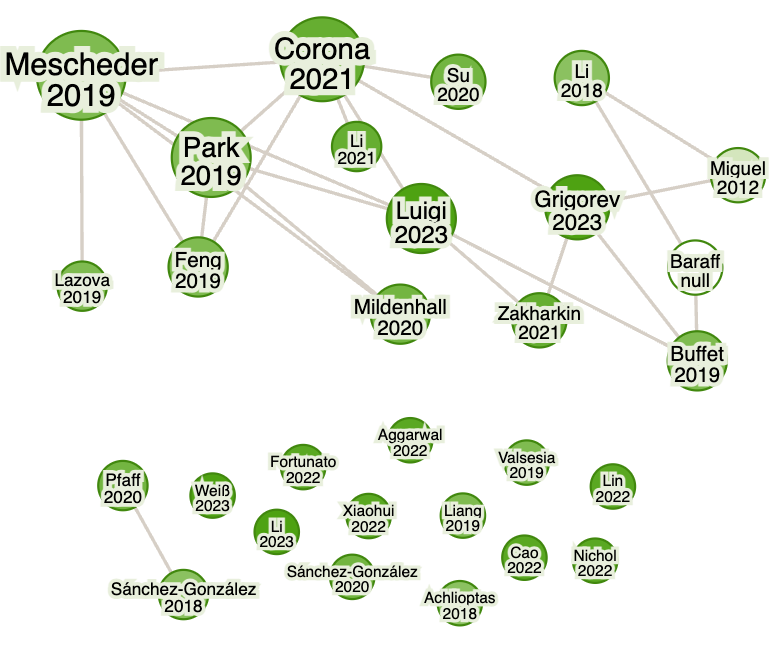}
  \caption{Graphical representation of methods}
  \label{fig:Figure}
\end{figure}

The study begins with an introduction of preliminaries necessary to gain a comprehensive understanding of the key concepts applied in most techniques of draping. Fig. \ref{fig:Figure} represents a graphical representation of the methods discussed in this research. The Physics-based simulations category encompasses an array of physics-based models, material behaviour, time integration and acceleration and differentiable simulation. Representations include 3D meshes, 3D point clouds and neural fields.

The importance of garment draping goes beyond the field of fashion and offers opportunities for innovation in many ways such as virtual try-on systems, virtual reality environments, and digital avatars in the gaming and entertainment sectors. This analysis can facilitate informed decision-making and future advancements in the field.

\section{Preliminary}

 Over the last decade, with the emergence and integration of new technology, a great amount of research has been dedicated to simulating cloth accurately \cite{stable_responsive_cloth, multiscale_meshGraphNets, deepCloth}. Cloth can demonstrate strong resistance to stretch and weak resistance to bending. The buckling of a thin material involves a very unstable state that is difficult to replicate. Therefore, cloth simulation techniques must be stable enough for practical implementation.

The term "time steps" refer to discrete intervals of time at which the state of the cloth is calculated by the simulation. Simulations are typically broken down into smaller and discrete steps to make the calculations more resource and time efficient.

Linear Blend Skinning (LBS) is a common technique used in computer graphics to animate 3D models in a computationally efficient way. The general idea is to deform the mesh to create the illusion of movement. In LBS, each vertex of the character is influenced by a set of "bones" in the skeleton. The movement of these bones is controlled by an animator, as a result, the vertices of the mesh are transformed accordingly, to simulate the movement.

A distance field \cite{distfield} is a representation where at each point, the distance from that point to the closest point on any object within the domain is known. A Signed Distance Field (SDF) defines the signed distance from any point in space to the nearest surface of an object whereas an Unsigned Distance Field (UDF) represents the unsigned distance. An SDF can be zero if it's exactly on the surface of the object, positive if the point is outside the object and negative if it's inside the object. On the other hand, UDF only represents the distance magnitude without indicating the relative position of the point with respect to the object.

Pose Space Deformation \cite{posespacedeform} is a character animation technique used to make deformations more expressive. A set of corrective deformations, called "poses", are precomputed which capture the desired deformations for specific poses of the character. These poses are then applied during animation to enhance the overall deformation quality.

Differentiable simulation \cite{diffeqn} is a computational technique where the simulation model is treated as a differentiable function. It allows for the use of gradient-based optimization methods to optimize the parameters of the simulation model.

Physics-based methods model the behavior of cloth using principles of physics \cite{complex_graphnets}. It simulates how cloth interacts with external forces like gravity, wind, and collisions with other objects. It involves representing the cloth as a collection of interconnected particles, with each particle subject to various forces such as gravity, friction, collision, etc.

Stiffness parameters \cite{stiffparams} are the numerical values that control the rigidity or flexibility of cloth material. It determines the resistance of the cloth to deformation and bending when it's under influence of various forces.

Strain limiting is the concept of imposing bounds on the membrane deformation of a material or structure under applied loads. It is however rare to find a method that can guarantee the enforcement of strain limits. This results in unpredictable and inconsistent differences in material behavior for each scene and step of the simulation \cite{cipc}. Frictional forces are the resistive forces that oppose the tendency of motion between two surfaces in contact. One of the biggest challenges in reproducing dynamic solid materials, such 3D soft bodies, cloth, or hair, is accurately accounting for contact and friction. At contact, objects should realistically dissipate energy, following the Coulomb law for dry friction \cite{dry_frictional_contact}. Future time steps also experience friction forces that add to the realism by producing stable folds and wrinkles \cite{robust_collision}.

Physics-based losses are energy losses that occur due to physical principles such as friction or resistance, leading to a decrease in the performance of a system. They can be fundamental to self-supervision techniques \cite{snug}. It can incorporate membrane strain energy of the deformed garment, bending energy, penalty for collisions and more \cite{drapenet}. Implicit methods are numerical techniques for solving differential equations, taking into account both the current and future states of a system to determine the system's future state. The architecture can be conditioned on the body pose and shape \cite{smplicit} for garment draping. Implicit fields are used which are scalar functions that map points in space to a single value. This value determines whether the point is inside or outside the surface.

\begin{figure}[htbp]
    \centering
    \includegraphics[width=0.9\linewidth]{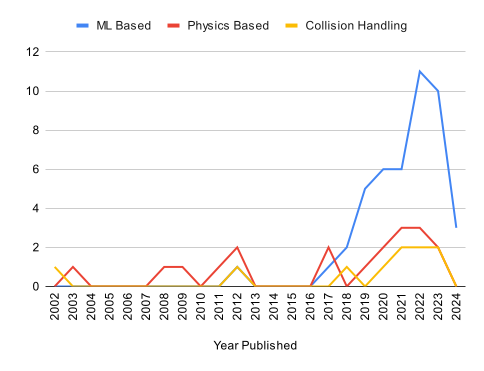}
    \caption{Category vs. Year Published}
    \label{fig:Line Graph}
\end{figure}

Techniques that use algorithms to help identify patterns in data and make predictions or decisions are referred to as machine learning-based methods. In machine learning, self-supervised techniques use models that learn representations from unlabeled data through tasks that allow the models to produce their own supervision signals. It is possible to learn complex tasks without requiring ground-truth data \cite{snug} which often leads to simpler solutions than a traditional simulator. Figure \ref{fig:Line Graph} represents the trends of machine learning based methods, physics based methods and collision handling techniques throughout the years.

\begin{table}
\caption{Classification of Methods}
\label{tab:example}
\centering 
\renewcommand{\arraystretch}{1.5} 
\begin{tabular}{|>{\centering\arraybackslash}p{4cm}|p{3cm}|p{7cm}|} %
\hline
\multicolumn{1}{|c|}{Classification} & \centering Type & Methods \\
 \hline
\multirow{7}{*}{\centering Representations}
& 3D Meshes & SMPL \cite{smpl}, HOOD \cite{hood}, Multiscale GNNs \cite{multiscale_graph_neural_networks}, Multiscale MeshGraphNets \cite{multiscale_meshGraphNets}, Graph Based Synthesis \cite{graph_based_synthesis}, GraphNet \cite{graphnet}, Complex GraphNets \cite{complex_graphnets} \\
\cline{2-3}
& 3D Point Clouds &  Pointe \cite{pointe}, Lion \cite{lion}, Graph 3D Point \cite{graph_3D_point}, Learning Representations \cite{learning_representations}, FITE \cite{point_based_implicit}, Point Based Modeling \cite{point_based_modeling} \\
\cline{2-3}

& Neural Fields & NeRF \cite{nerf}, Occupancy Networks \cite{occupancy_networks}, Implicit Fields \cite{implicit_fields}, DeepSDF \cite{deepsdf}, Texture 360 \cite{texture_360}, Icon \cite{icon}, Photorealistic Monocular \cite{photorealistic_monocular}, Neural Gif \cite{neural_gif}, Meta Avatar \cite{meta_avatar} \\
\cline{2-3}

\hline
\multirow{2}{*}{\centering Physics Based Methods} 
& Material Behaviour & Estimating Cloth Simulation Parameters \cite{cloth_params_from_vid}, Data-Driven Estimation \cite{data_driven_estimation}, Data Driven Elastic Models \cite{data_driven_elastic_model}, Dry Frictional Contact \cite{dry_frictional_contact}, How Will it Drape Like \cite{how_will_it_drape}, Estimation for Woven Fabric \cite{clyde_et_al}, Cloth Recovery from Video \cite{learning_based_recovery}, Drape Estimation for Non Linear Stiffness\cite{non_linear_stiffness}, Measuring Static Friction Coefficient\cite{learning_to_measure}, Estimating Cloth-Body and Cloth-Cloth Friction\cite{visual_approach} \\
\cline{2-3}

& Time Integration and Acceleration & Large Steps \cite{large_steps_in_clothsim}, Asynchronous \cite{asynchronous_cloth}, GPU Based \cite{gpu_based}, Adaptive Remeshing \cite{adaptive_anis_remesh} \\
\cline{2-3}

& Differentiable Simulation & DiffCloth \cite{diffcloth}, Differentiable Cloth for Inverse Problems \cite{diffcloth_inverse_problems}, Estimating Cloth Parameters \cite{estimating_cloth_params_with_position}, Learning Material Parameters \cite{diff_material_robotic_fish} \\
\cline{2-3} 

\hline
\multirow{2}{*}{\centering ML Based Methods} 
& Learned Deformation Models & DRAPE \cite{drape}, TailorNet \cite{tailornet}, Animation using Bone Motion Networks \cite{bone_motion_networks}, DIG \cite{dig}, SNUG \cite{snug}, PHORUM \cite{photorealistic_monocular}, AniDress \cite{anidress}, DLCA-Recon \cite{dlca}, Self-Supervised Collision Handling \cite{self_supervised_collision_vto}, PBNS \cite{pbns}, Learning Based Animation \cite{vto}, GarNet \cite{garnet}, CLOTH3D \cite{cloth3d}, Motion Guided Network \cite{motion_guided}, Multi-Res Pyramid \cite{multi_res_pyramid}, DeePSD \cite{deepsd}, GenSim \cite{gensim}, GarSim \cite{garsim}, Garment3DGen \cite{garment3dgen}, Expression Based Wrinkles \cite{mesh_tension}, Multi-Garment Net \cite{multi_garment_net}\\
\cline{2-3}

& Implicit Representation & ISP \cite{isp}, Multi-Res Pyramid \cite{multi_res_pyramid}, DeepCloth \cite{deepCloth}, I-Cloth \cite{icloth}, Layered Garment Net \cite{layered_garment_net}, SMPlicit \cite{smplicit}, Robust SDF \cite{robust_sdf}, DIG \cite{dig}, Layered Garment Net \cite{layered_garment_net}, FITE \cite{point_based_implicit}, Implicit Untangling \cite{implicit_untangling}, SNARF \cite{snarf}, gDNA \cite{gDNA} \\

\cline{2-3}
& GNNs & Multiscale MeshGraphNets \cite{multiscale_meshGraphNets}, Complex GraphNets \cite{complex_graphnets}, Graph Based Synthesis \cite{graph_based_synthesis}, Multiscale GNNs \cite{multiscale_graph_neural_networks}, NCloth \cite{NCloth}, GNN Physics Engine \cite{gnn_physics_engine}, GraphNet \cite{graphnet}, HOOD \cite{hood}, Bi-Stride \cite{bistride} \\
\cline{2-3} 

& Training Methods & Cloth3D \cite{cloth3d}, Tex2Shape \cite{tex2shape}, Deep Garment \cite{deep_garment}, AvatarCLIP \cite{avatarCLIP}, USR \cite{usr}, DRAPE \cite{drape}, SNARF \cite{snarf} \\
\cline{2-3}

\hline
Collision Handling
& \multicolumn{2}{p{10cm}|}{Robust Collision \cite{robust_collision}, Contact Mechanics \cite{async_contact_mechanics}, IPC \cite{ipc}, C-IPS \cite{cipc}, I-Cloth \cite{icloth}, Self-Supervised Collision \cite{self_supervised_collision_vto}, SNUG \cite{snug}, ClothCombo \cite{clothcombo}, ULNeF \cite{unlef}} \\
\hline
\end{tabular}
\end{table}

\begin{figure}[htbp]
    \centering
    \includegraphics[width=0.80\textwidth]{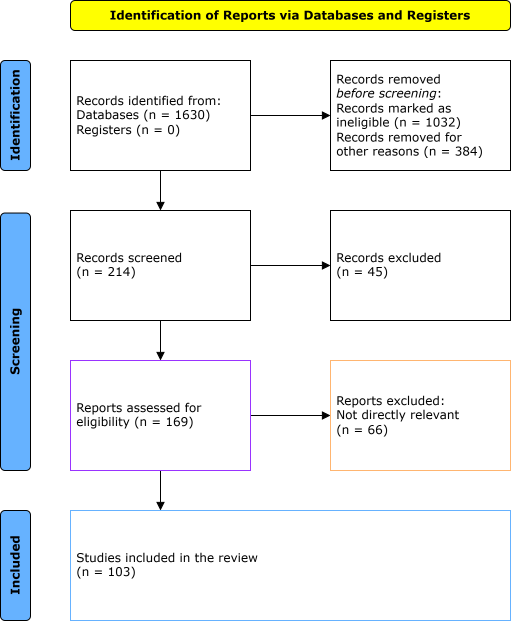} 
    \caption{PRISMA flow diagram for this comparative study}
    \label{fig:PRISMA}
\end{figure}

Strong numerical integration techniques are supported by mesh representations, and their resolution can be adjusted to achieve advantageous efficiency and accuracy trade-offs \cite{graphnet}. A computational method for simulating the dynamic behaviour of deformable objects in simulations is Projective Dynamics. It was introduced as an efficient method to yield a stable solution to the dynamics of nodal systems subject to stiff internal forces \cite{dry_frictional_contact}. Figure \ref{fig:PRISMA} demonstrates the PRISMA flow diagram for this comparative study. 

AMASS \cite{amass} is a large database of human motion that consolidates different optical marker-based motion capture datasets by representing them within a common framework. AMASS is used for animation, visualization, and generating training data for deep learning. Cloth3D \cite{cloth3d} is the first big scale synthetic dataset of 3D clothed human sequences that contains various types of garment and their properties. Clothes
are simulated on top of thousands of different pose sequences and body shapes. Skinned Multi-Person Linear model (SMPL) \cite{smpl} is a skinned vertex-based model that represents different kinds of body shapes in natural human poses. Table \ref{tab:example} shows an overview of classification of various methods.

\section{Representations}
\subsection{3D Meshes}

Various 3D representation techniques are used to represent garments to be processed for computer graphics tasks. One of the most widely used representations are polygons or 3D meshes. They are made up of a collection of faces (usually triangles or quads), edges, and vertices that define the surface shape of the object. Because of its effectiveness and simplicity, polygonal meshes are useful and commonly used in computer graphics. Methods like SMPL \cite{smpl} which are used by a large number of draping methods \cite{snug, hood} to represent the body mesh, learn from a large 3D dataset to represent  different kinds of body shapes in natural human poses.

\begin{figure}[h]
  \centering
  \begin{subfigure}{0.3\textwidth}
    \centering
    \includegraphics[width=\linewidth]{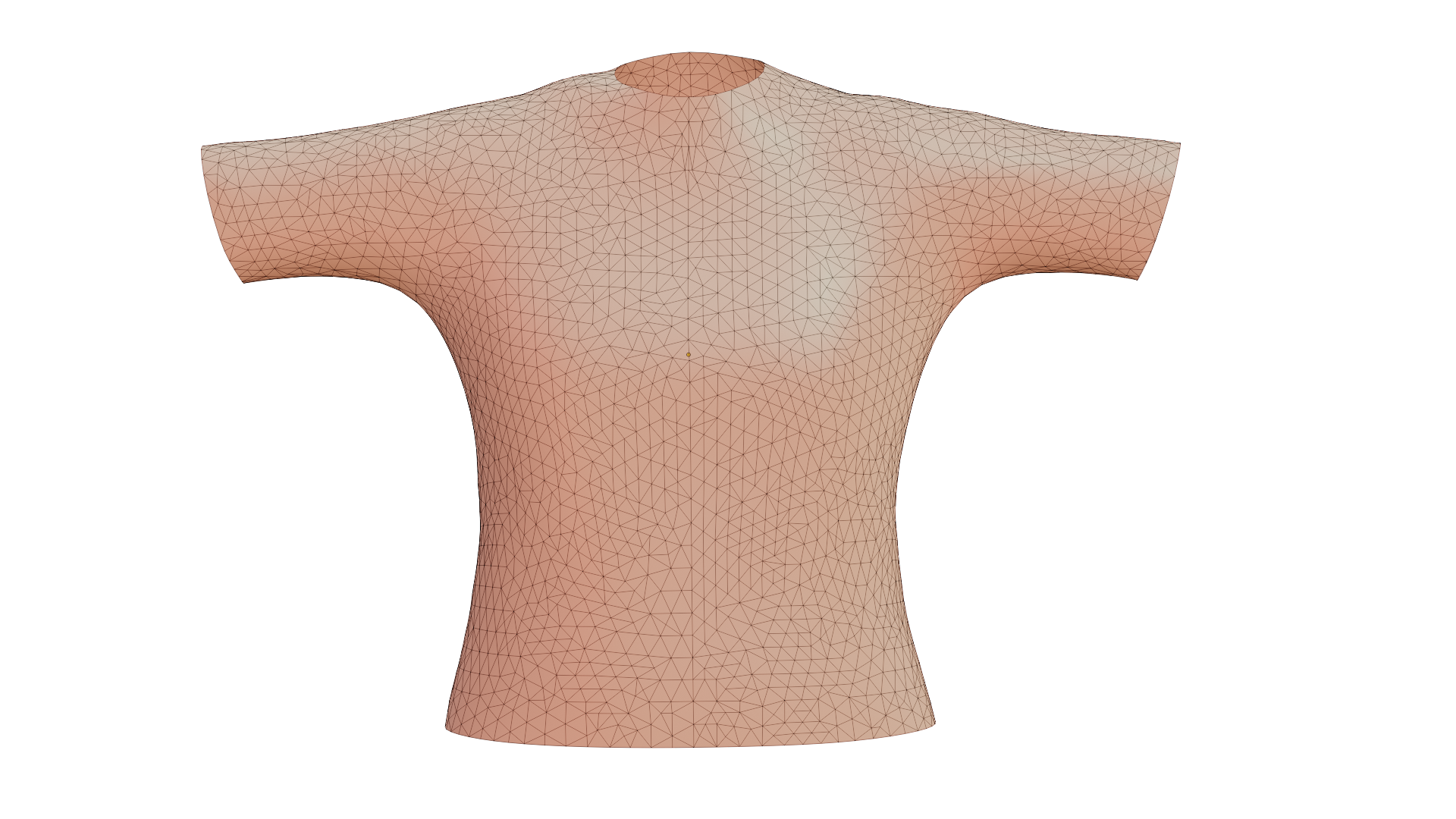}
    \caption{3D Meshes}
    \label{fig:Mesh}
  \end{subfigure}
  \begin{subfigure}{0.3\textwidth}
    \centering
    \includegraphics[width=\linewidth]{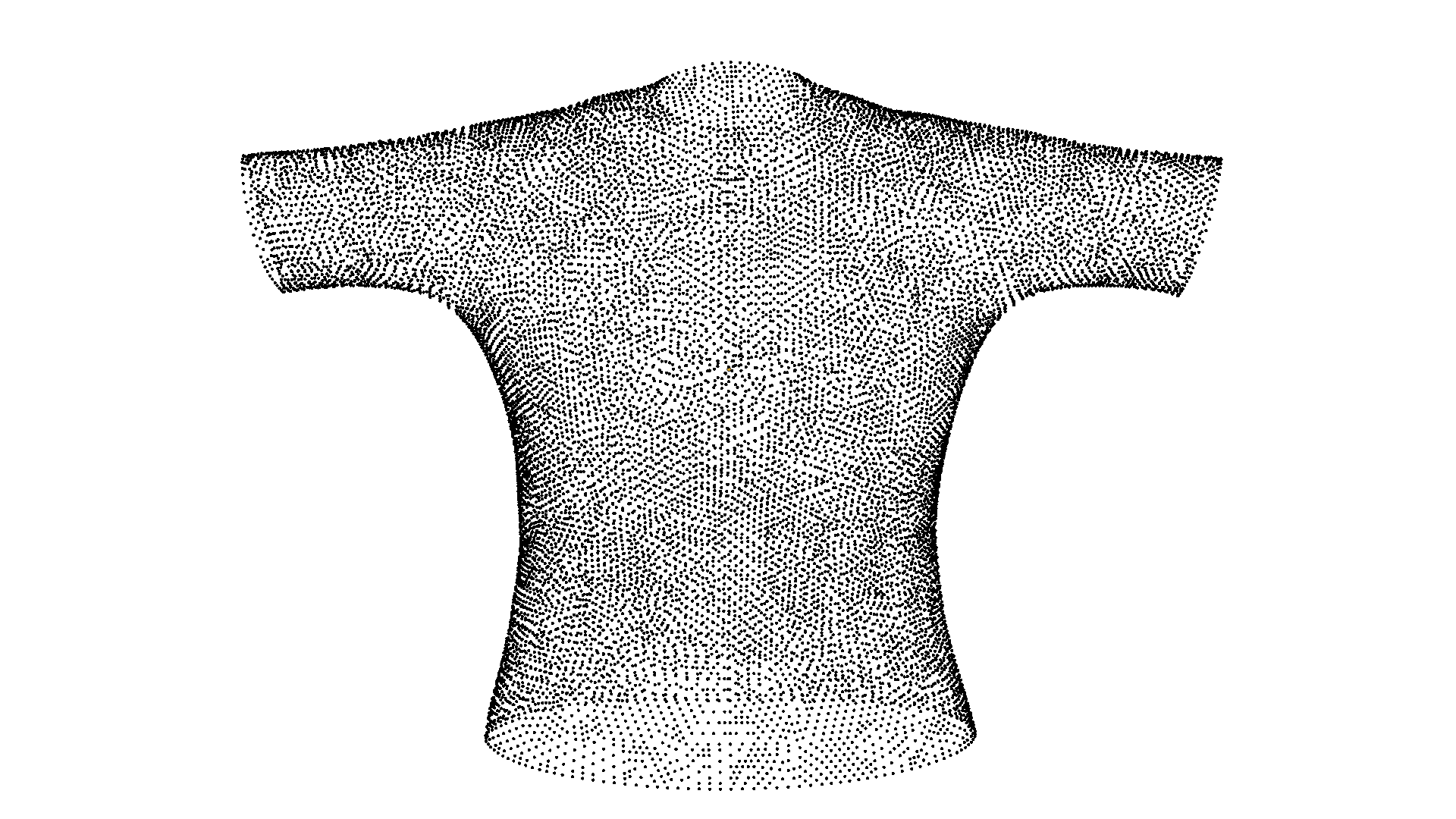}
    \caption{3D Point Clouds}
    \label{fig:PC}
  \end{subfigure}
  \begin{subfigure}{0.3\textwidth}
    \centering
    \includegraphics[width=\linewidth]{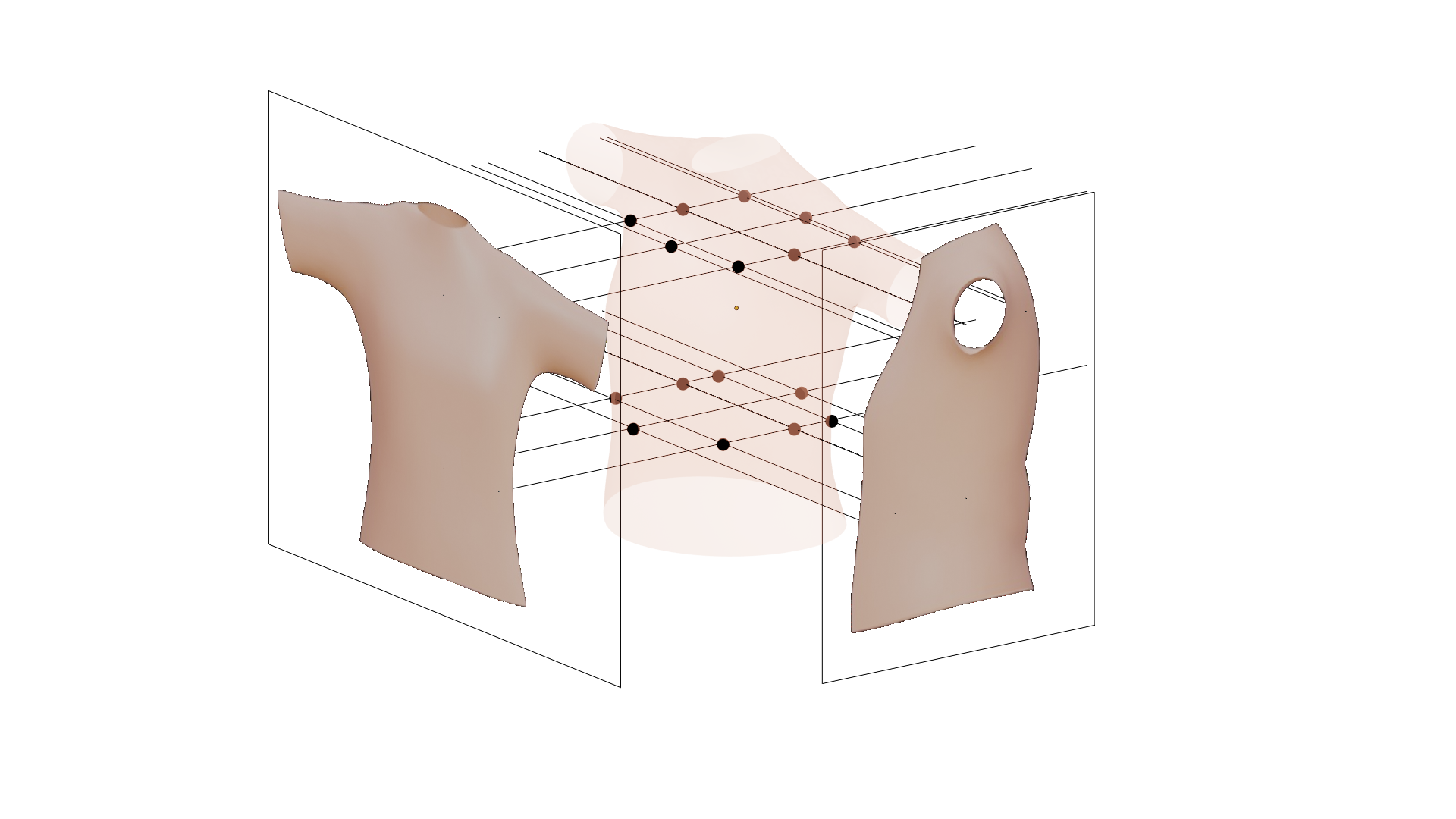}
    \caption{Neural Fields}
    \label{fig:NERF}
  \end{subfigure}
  \caption{Representations}
  \label{fig:Representations}
\end{figure}

The main issues with mesh data is their irregularity which prevents them from being commonly utilised in deep learning techniques. Hence several methods \cite{snug, drape, garment3dgen, tailornet, drapenet} make use of underlying garment templates which are limited to the specific templates they were trained on. But recent methods \cite{hood, multiscale_graph_neural_networks, multiscale_meshGraphNets, graph_based_synthesis, graphnet, complex_graphnets} overcome these complexities by using Graph Neural Networks to leverage underlying mesh representations and produce faster and more accurate simulations. Figure \ref{fig:Mesh} visualises the 3D Mesh of a t-shirt.

\subsection{3D Point Clouds}
Point clouds are collections of three-dimensional points in space with each point representing a feature or surface point on the garment. Point clouds are frequently generated by 3D scanning methods like photogrammetry or LiDAR. This type of representation is used by several of the popular 3D generation algorithms \cite{pointe, lion, graph_3D_point, learning_representations} and also by various garment generation and processing algorithms \cite{point_based_implicit, point_based_modeling}. This is due to the ease of capture and representation of 3D data by scanning instruments, however processing them can be challenging because of the lack of connectivity information. Figure \ref{fig:PC} visualises the 3D point clouds of a t-shirt.

\subsection{Neural Fields}
Implicit surfaces \cite{real_time_radiance_field} represent objects as a set of a scalar function defined over the 3D space. Within the object, the function evaluates to positive values (Signed distance function) \cite{robust_sdf}, and outside, to negative values or zero (Unsigned distance function). Unlike meshes, or point clouds, deep implicit functions \cite{nerf, occupancy_networks, implicit_fields, deepsdf} can represent garments without limits to resolution or topology. Saito et al. \cite{texture_360} introduced reconstruction of 3D humans from images. Other works such as \cite{icon, photorealistic_monocular, neural_gif, meta_avatar} significantly improve and speed up the performance of 3D reconstruction. Although these methods are capable of capturing varied topology, they are expensive to render, require large amounts of training data, and are incompatible with current graphics pipelines. Figure \ref{fig:NERF} visualises the neural fields of a t-shirt.

\begin{figure}[htbp]
    \centering
    \includegraphics[width=1\textwidth]{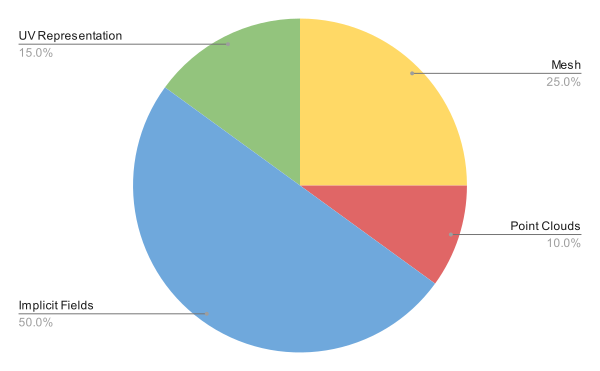} 
    \caption{Distribution of 3D Representation techniques}
    \label{fig:Rep}
\end{figure}

Figure \ref{fig:Rep} represents the distribution of 3D Representation techniques in this study.

\section{Physics Based Methods}
Cloth simulation and draping techniques have been a vastly explored topic in graphics. Traditionally physics based modelling methods \cite{stable_responsive_cloth, discrete_shells, non_linear_stiffness, robust_collision, complex_wrinkle} were employed to achieve realistic draping results at high computational cost. Newer methods like CIPC \cite{cipc}, IPC \cite{ipc} and other yarn based methods \cite{yarn_level_sim, yarn_level_with_persistent_contacts, simulating_knitted_cloth} produce realistic results in considerably lesser computation time.

\subsection{Material Behaviour}
Material parameters are greatly responsible for the look of a particular fabric, which can be difficult to tune and adjust for realistic cloth simulations. As described in \cite{cloth_params_from_vid, learning_based_recovery}, real fabric video data can be used to estimate the simulation data and can further be tested by a perceptually motivated metric. Estimation of material parameters from real fabric footage has also been investigated in recent years. Rasheed et al. \cite{visual_approach} and \cite{learning_to_measure} use videos with a causal capture setup whereas Rodríguez-Pardo et al. \cite{how_will_it_drape} identify material parameters from just two multiview images.

Many of the current techniques use models with manually selected stiffness parameters which have the drawback of inaccuracy of materials to their real-world counterparts. Data-driven techniques can prove to be more accurate as shown in \cite{data_driven_estimation, clyde_et_al, data_driven_elastic_model, drape_sim}, where fitting is accomplished using non-linear models. 

\subsection{Time Integration and Acceleration}
To avoid numerical instability, most cloth simulation systems require small time steps. David Baraff and Andrew Witkin presented a new technique \cite{large_steps_in_clothsim} for enforcing constraints on individual cloth particles with an implicit integration method with large time steps. The resulting simulation system is significantly faster than previous accounts of cloth simulation systems in the literature. 

Asynchronous variants of collision handling and time integration can address limitations of implicit integration methods, such as over-damped simulations and suppressing details \cite{asynchronous_cloth}. It achieves superior quality while maintaining comparable computation times. Recently, the use of GPUs \cite{gpu_based} and other methods to accelerate physics based simulations are being investigated. 

\subsection{Differentiable Simulation}
Differentiable physics simulation is a technique that uses gradient-based methods to learn and control physical systems. It has been successful in rigid-body systems, fluidic systems, and deformable-body systems. Recent work introduces a differentiable cloth simulator \cite{diffcloth} with additional gradient information. The simulator's efficacy is demonstrated in various applications, including system identification of frictional coefficients in cloth simulation, inverse garment design for computer animation, and motion planning of robotic manipulators in robot-assisted dressing.

Differentiable Cloth Simulation for Inverse Problems \cite{diffcloth_inverse_problems} provides a unified approach to various inverse problems involving cloth. This reduces gradient computation to a linear system, enabling fast simulation and backpropagation. It has potential in applications like physical parameter estimation and motion control of cloth. Differentiable methods have also been used to estimate material parameters \cite{estimating_cloth_params_with_position} and \cite{diff_material_robotic_fish}.

\section{Machine Learning Based Methods}

To overcome the limitations of physics based models, neural networks in combination with linear blend skinning are used to learn the deformations of garments for draping and animation tasks. Machine Learning based methods can generate visually appealing results in a fraction of the time \cite{snug, tailornet}. Using implicit representations \cite{smplicit, isp, implicit_untangling} allow for working with varied garment representation and resolution quality. It simultaneously ensures collision free simulations as compared to traditional physics based methods.

\subsection{Learned Deformation Models}
To overcome the limitations of physics based models, neural networks in combination with linear blend skinning are used to learn the deformations of garments using shape and pose parameters of the body \cite{drape, vto, tailornet, dig, snug, learning_generative_clothing}.

Such methods \cite{tailornet, drape, snug} yield plausible results, however, they do not work well with loose-fitted garments such as long skirts and dresses due to their reliance on skinning. To overcome this, several alternative methods were devised. HOOD\cite{hood} uses Graph Neural Networks, \cite{self_supervised_collision_vto, robust_skin_weights} extend skinning weights to garment points beyond the body, while others \cite{bone_motion_networks, anidress} use Bone-Driven Networks.

Another set of methods \cite{photorealistic_monocular, dlca, 3d_reconstruction_of_interacting, avatargen, multi_garment_net, cloning_outfits}, use implicit neural representations to reconstruct garments from RGB images or videos. A Deep neural network named PHORUM \cite{photorealistic_monocular} generates photorealistic 3D human reconstruction using a monocular RGB image. It addresses limitations in reshaping geometry, and illumination effects. DLCA-Recon \cite{dlca}, on the other hand, proposes a dynamic deformation field with SMPL weight propagation to optimize the deformations.

Although state-of-the-art methods exist for garment draping \cite{drape, snug, bone_motion_networks, self_supervised_collision_vto}, they are garment specific and therefore can only estimate deformations for the garments they were initially trained on. Such models require retraining for new garments which proves to be inefficient and time consuming. Some methods \cite{cloth3d, point_based_modeling} overcome this and aim to learn garment spaces from a large dataset of 3D clothing. Others \cite{isp, multi_res_pyramid, deepCloth} use representations in the texture space using UV maps. DeepCloth \cite{deepCloth} enables seamless transition of garments between various shapes. It first represents the garments using topology UV maps which are used to reconstruct the garments. These reconstructed garments are then animated based on style and shape allowing them to transition from one configuration to another. Recently, implicit models such as \cite{icloth, layered_garment_net, isp, deepCloth} have also been used due to their flexibility and ability to represent a wide range of garments. 

\subsection{Implicit Representation}
Current learning-based methods necessitate training distinct models for various types of clothing, a study SMPLicit \cite{smplicit} presents a new generative model that simultaneously captures body position, form, and apparel geometry that can be used for a wide range of clothing items. SMPLicit's representation flexibility is based on an implicit model that is influenced by SMPL \cite{smpl} human body parameters and a trainable latent space that is semantically interpretable and corresponds with clothing attributes. It can be used to fit 3D scans and reconstruct clothed humans in images that pose challenges, such as multi-layered garments or strong body occlusions due to the presence of multiple people.

Representing collisions involving simulated cloth, rigid bodies, and deformable solids can be complex, where Signed distance fields (SDFs) are generally used. Research on SDFs \cite{robust_sdf} introduces a per-element local optimisation method to determine the nearest positions between the SDF isosurface and mesh elements. This enables the creation of precise contact sites between sharp point-face pairings and manage smoothly changing edge-edge contact.

In \cite{dig}, the authors propose an end-to-end differentiable pipeline for garment draping that uses implicit surfaces and learns a skinning field based on body model parameters. The method limits interpenetrations and artifacts, leading to more accurate garment reconstructions and deformations. Layered-Garment Net (LGN) \cite{layered_garment_net} is the first research work that generates intersection-free multiple layers of garments from a person's near front-view image. 

First-Implicit-Then-Explicit (FITE) \cite{point_based_implicit} is a framework for modeling human avatars in clothing. It learns implicit surface templates and uses them to generate point sets for capturing pose-dependent deformations like wrinkles and proposes diffused skinning for loose clothing. Implicit Untangling \cite{implicit_untangling} presents a method for untangling cloth layers, allowing for collision-free animation. It uses an intermediate, implicit representation to generate plausible static shapes of garments with safe animations of complex layered clothing. Chen et al. \cite{snarf} introduces a forward skinning model named SNARF that combines the advantages of linear blend skinning for polygonal meshes with neural implicit surfaces, generalizing better to unseen poses while preserving accuracy. gDNA \cite{gDNA} proposes a novel method for generating 3D human avatars with varied identities and shapes in various garments using a multi-subject forward skinning module, learning from a few posed scans per subject.

\subsection{Graph Neural Networks (GNN)}
3D garments can best be represented using graphs with nodes and edges. GNNs \cite{multiscale_meshGraphNets, complex_graphnets,graph_based_synthesis, multiscale_graph_neural_networks, NCloth, gnn_physics_engine}, have shown promising results in replacing physics-based simulation due to their speed and ability to utilise the existing mesh representations of physical objects. They utilise predictive algorithms to estimate accelerations for each individual node of the garment. The work in \cite{graphnet, complex_graphnets}, uses message passing networks to learn dynamics of physical systems from mesh-based simuations. However, it is difficult to set the number of message passing steps in advance which results in undesirable results with high resolution simulations.

Similar to traditional remeshing techniques \cite{adaptive_anis_remesh}, various newer methods such as \cite{hood, multiscale_graph_neural_networks, multiscale_meshGraphNets, bistride} use a hierarchial approach to  predict accurate dynamics of a high-resolution system on a much coarser mesh. Artur Grigorev et al. in \cite{hood} utilise a hierarchical structure to perform local as well as global changes by adding coarse edges to the same graph whereas \cite{multiscale_meshGraphNets, multiscale_graph_neural_networks} pass messages on different resolutions (fine and coarse) significantly improving the accuracy of MeshGraphNets \cite{graphnet}. Further, HOOD \cite{hood} also operates in a self supervised manner by optimising a physics based loss function. Each transformation results in a set of messages that are then used to update feature vectors. 

\subsection{Training Methods}
Data-driven approaches have proven to be useful in learning garment deformations as a function of human body. Methods such as \cite{vto, drape, snarf, multi_res_pyramid, clothcap} employ a large dataset and use supervised learning to obtain satisfactory results in terms of garment deformations. 

The training data for such data-driven techniques can be obtained from either physics-based simulations \cite{cloth3d} or 3D reconstruction methods based on multi-view inputs \cite{tex2shape, deep_garment, avatarCLIP, usr}. Text2Shape \cite{tex2shape} is a method that infers detailed human body shapes from a single photograph, including face, hair, and clothing. The model uses a partial texture map and estimates detailed normal and vector displacement maps. DRAPE \cite{drape} uses the technique of factoring to animate clothing on various shapes and poses without manual intervention. However, such methods require large amounts of data, to tackle this problem, SNARF \cite{snarf} defines a forward skinning model that uses implicit differentiation to learn forward deformations without need for direct supervision while training. USR \cite{usr} proposes an unsupervised separated 3D garments and human reconstruction model to improve the reconstruction of dressed people from images.

Machine Learning Methods can also be combined with physics based losses. Several state-of-the-art unsupervised methods \cite{snug, hood, pbns, drapenet, lee_and_lee} have been developed that are capable of producing similar or better results without needing large amounts of data. HOOD \cite{hood} relies on message passing mechanism of graph-based neural networks in order to model coarse and fine deformations. Whereas, SNUG \cite{snug} treats the update to vertex position as an optimization problem which is resolved using backward Euler equation. This modification allows the network to be trained in an unsupervised way without the need for any ground truth captions.

\section{Collision Handling}
Collisions are a considerable bottleneck in cloth simulation. A collision handling algorithm \cite{robust_collision} may be used with any internal dynamics simulation technique such as shearing, stretching, and bending to produce interference-free and robust complicated motion. It combines a fast repulsion force with a geometric collision method to model both static and kinetic friction with cloth thickness.
Through a combination of asynchronous variational integrators, kinetic data structures, and a discretization of the contact barrier potential, \cite{async_contact_mechanics} can simulate complex contact scenarios. It handles problems regarding sharp features and sliding.

Incremental Potential Contact (IPC) \cite{ipc} is a new model and algorithm for variationally solving implicitly time-stepped nonlinear elastodynamics. It represents the intricate interplay of deforming solids in contact. IPC \cite{ipc} offers an adaptable, effective, and completely practical solution for mesh-based volumetric nonlinear elasticity simulations. Codimensional Incremental Potential Contact (C-IPC) \cite{cipc} extends IPC \cite{ipc} from volumetric deformables to codimensional and mixed-dimensional structures that are accurately frictionally behaved, have controlled thickness, fully connected strain limiting, and are non-intersecting at all time steps .

I-Cloth \cite{icloth} integrates Incremental Continuous Collision Detection using Spatial Hashing and GPU-based Non-Linear Impact Zone Solver to perform collision handling. One of the first algorithms to learn nearly collision-free deformations of clothing is \cite{self_supervised_collision_vto}. It achieves this by designing subspace for the garment that produces a differentiable, canonical configuration, along with a diffused, volumetric representation of the underlying body. Bone-Driven Motion Networks \cite{bone_motion_networks, anidress} have been used to introduce the first learning-based technique to learn the intricate deformations. For simplicity, it employs LBS as the skinning mechanism. This methodology more accurately predicts complicated deformations that are commonly seen in loose-fitting clothing when compared to systems that solely use 3D coordinates. 

Collisions between body and garment or between numerous garments pose a challenge to realistically draping garments over human body. Igor Santesteban et al. \cite{snug} define a collision penalty term while computing static energy in order to enforce the garment follow the motion of the body. This enables the model to replicate complex behaviours including wrinkles at various scales.

Numerous approaches for draping garments on the human models are realistic and fast yet they do not handle multiple layers of clothing. ISP \cite{isp} is a combination based technique that defines the 2D shape of the garment using Signed Distance Function and 3D shape by mapping 2D to 3D. ClothCombo \cite{clothcombo} solves this problem using a GNN-based network that first creates feature embedding for each cloth. Then, the draping network deforms the clothes to fit the target body while the untangling network resolves collisions between layers or clothes.

Untangled Layered Neural Fields for Mix-and-Match Virtual Try-On (ULNeF) \cite{unlef} explores the use of neural fields for mix-and-match virtual try-on problems. It proposes a neural model that focuses on the interaction between layered neural fields. 

\section{Conclusion}

In this review paper, we provide an extensive analysis of methods used in 3D fashion design, virtual try-ons, and animations for multilayered garment draping. Through a meticulous comparative study encompassing various methodologies, including diverse models, material behavior simulations, collision handling strategies, and the integration of graph neural networks, we have provided a comprehensive study of the domain. It equips researchers with a baseline of necessary insights to make informed decisions when selecting the most suitable approach for their specific needs. These techniques aim to accurately portray the complexities of garment deformations and small wrinkles in digital simulations. Our work aims to be a useful tool for researchers, artists, and programmers who want to improve their comprehension and application of dynamic multilayered 3D cloth draping visualisations.

\section{Acknowledgement}

\bibliographystyle{splncs04}
\bibliography{sn-bibliography}

\begin{thebibliography}{100}
\providecommand{\url}[1]{\texttt{#1}}
\providecommand{\urlprefix}{URL }
\providecommand{\doi}[1]{https://doi.org/#1}

\bibitem{robust_skin_weights}
Abdrashitov, R., Raichstat, K., Monsen, J., Hill, D.: Robust skin weights transfer via weight inpainting. In: SIGGRAPH Asia 2023 Technical Communications. SA '23, Association for Computing Machinery, New York, NY, USA (2023). \doi{10.1145/3610543.3626180}

\bibitem{learning_representations}
Achlioptas, P., Achlioptas, P., Diamanti, O., Diamanti, O., Mitliagkas, I., Mitliagkas, I., Guibas, L.J., Guibas, L.J.: Learning representations and generative models for 3d point clouds. International Conference on Machine Learning  (2018). \doi{null}

\bibitem{layered_garment_net}
Aggarwal, A., Aggarwal, A., Wang, J., Wang, J., Hogue, S., Hogue, S., Ni, S., Ni, S., Budagavi, M., Budagavi, M., Guo, X., Gu, X.: Layered-garment net: Generating multiple implicit garment layers from a single image. Cornell University - arXiv  (2022). \doi{10.48550/arxiv.2211.11931}

\bibitem{tex2shape}
Alldieck, T., Pons-Moll, G., Theobalt, C., Magnor, M.: Tex2shape: Detailed full human body geometry from a single image  (2019)

\bibitem{photorealistic_monocular}
Alldieck, T., Zanfir, M., Sminchisescu, C.: Photorealistic monocular 3d reconstruction of humans wearing clothing  (2022)

\bibitem{large_steps_in_clothsim}
Baraff, D., Witkin, A.: Large Steps in Cloth Simulation. Association for Computing Machinery, New York, NY, USA, 1 edn. (2023)

\bibitem{cloth3d}
Bertiche, H., Madadi, M., Escalera, S.: Cloth3d: Clothed 3d humans  (2020)

\bibitem{pbns}
Bertiche, H., Madadi, M., Escalera, S.: Pbns: Physically based neural simulator for unsupervised garment pose space deformation  (2021)

\bibitem{deepsd}
Bertiche, H., Madadi, M., Tylson, E., Escalera, S.: Deepsd: Automatic deep skinning and pose space deformation for 3d garment animation. In: 2021 IEEE/CVF International Conference on Computer Vision (ICCV). pp. 5451--5460 (2021). \doi{10.1109/ICCV48922.2021.00542}

\bibitem{cloth_params_from_vid}
Bhat, K.S., Twigg, C.D., Hodgins, J.K., Khosla, P.K., Popovi\'{c}, Z., Seitz, S.M.: Estimating cloth simulation parameters from video. In: Proceedings of the 2003 ACM SIGGRAPH/Eurographics Symposium on Computer Animation. p. 37–51. SCA '03, Eurographics Association, Goslar, DEU (2003)

\bibitem{multi_garment_net}
Bhatnagar, B.L., Tiwari, G., Theobalt, C., Pons-Moll, G.: Multi-garment net: Learning to dress 3d people from images  (2019)

\bibitem{robust_collision}
Bridson, R., Fedkiw, R., Anderson, J.: Robust treatment of collisions, contact and friction for cloth animation. ACM Trans. Graph.  \textbf{21}(3),  594–603 (jul 2002). \doi{10.1145/566654.566623}

\bibitem{implicit_untangling}
Buffet, T., Rohmer, D., Barthe, L., Boissieux, L., Cani, M.P.: Implicit untangling: a robust solution for modeling layered clothing. ACM Trans. Graph.  \textbf{38}(4) (jul 2019). \doi{10.1145/3306346.3323010}

\bibitem{bistride}
Cao, Y., Cao, Y., Chai, M., Chai, M., Li, M., Li, M., Jiang, C., Jiang, C.: Bi-stride multi-scale graph neural network for mesh-based physical simulation. arXiv.org  (2022). \doi{10.48550/arxiv.2210.02573}

\bibitem{3d_reconstruction_of_interacting}
Cha, J., Lee, H., Kim, J., Truong, N.N.B., Yoon, J.S., Baek, S.: 3d reconstruction of interacting multi-person in clothing from a single image (2024)

\bibitem{anidress}
Chen, B., Shen, Y., Shuai, Q., Zhou, X., Zhou, K., Zheng, Y.: Anidress: Animatable loose-dressed avatar from sparse views using garment rigging model  (2024)

\bibitem{gDNA}
Chen, X., Jiang, T., Song, J., Yang, J., Black, M.J., Geiger, A., Hilliges, O.: gdna: Towards generative detailed neural avatars  (2022)

\bibitem{snarf}
Chen, X., Zheng, Y., Black, M.J., Hilliges, O., Geiger, A.: Snarf: Differentiable forward skinning for animating non-rigid neural implicit shapes. 2021 IEEE/CVF International Conference on Computer Vision (ICCV) pp. 11574--11584 (2021)

\bibitem{complex_wrinkle}
Chen, Z., Kaufman, D., Skouras, M., Vouga, E.: Complex wrinkle field evolution. ACM Trans. Graph.  \textbf{42}(4) (jul 2023). \doi{10.1145/3592397}

\bibitem{stable_responsive_cloth}
Choi, K.J., Ko, H.S.: Stable but responsive cloth. ACM Trans. Graph.  \textbf{21}(3),  604–611 (jul 2002). \doi{10.1145/566654.566624}

\bibitem{yarn_level_sim}
Cirio, G., Lopez-Moreno, J., Miraut, D., Otaduy, M.A.: Yarn-level simulation of woven cloth. ACM Trans. Graph.  \textbf{33}(6) (nov 2014). \doi{10.1145/2661229.2661279}

\bibitem{yarn_level_with_persistent_contacts}
Cirio, G., Lopez-Moreno, J., Otaduy, M.A.: Yarn-level cloth simulation with sliding persistent contacts. IEEE transactions on visualization and computer graphics  \textbf{23}(2),  1152--1162 (2016)

\bibitem{clyde_et_al}
Clyde, D., Teran, J., Tamstorf, R.: Modeling and data-driven parameter estimation for woven fabrics. In: Proceedings of the ACM SIGGRAPH / Eurographics Symposium on Computer Animation. SCA '17, Association for Computing Machinery, New York, NY, USA (2017). \doi{10.1145/3099564.3099577}

\bibitem{smplicit}
Corona, E., Pumarola, A., Aleny{\`a}, G., Pons-Moll, G., Moreno-Noguer, F.: Smplicit: Topology-aware generative model for clothed people. 2021 IEEE/CVF Conference on Computer Vision and Pattern Recognition (CVPR) pp. 11870--11880 (2021)

\bibitem{deep_garment}
Dan\'{z}\v{r}ek, R., Dibra, E., \"{O}ztireli, C., Ziegler, R., Gross, M.: Deepgarment: 3d garment shape estimation from a single image. Comput. Graph. Forum  \textbf{36}(2),  269–280 (may 2017). \doi{10.1111/cgf.13125}

\bibitem{diffeqn}
Degrave, J., Hermans, M., Dambre, J., Wyffels, F.: A differentiable physics engine for deep learning in robotics. Frontiers in Neurorobotics  \textbf{13} (2016)

\bibitem{distfield}
Desbrun, M., Meyer, M., Schroder, P., Barr, A.H.: Implicit fairing of irregular meshes using diffusion and curvature flow. In: Seminal Graphics Papers: Pushing the Boundaries, Volume 2. Association for Computing Machinery, New York, NY, USA (2023)

\bibitem{implicit_fields}
Feng, Y., Chen, Z., Zhang, H., Zhang, H.: Learning implicit fields for generative shape modeling. Computer Vision and Pattern Recognition  (2019). \doi{10.1109/cvpr.2019.00609}

\bibitem{multiscale_meshGraphNets}
Fortunato, M., Pfaff, T., Wirnsberger, P., Pritzel, A., Battaglia, P.W.: Multiscale meshgraphnets. ArXiv  \textbf{abs/2210.00612} (2022)

\bibitem{hood}
Grigorev, A., Black, M.J., Hilliges, O.: Hood: Hierarchical graphs for generalized modelling of clothing dynamics. In: 2023 IEEE/CVF Conference on Computer Vision and Pattern Recognition (CVPR). pp. 16965--16974 (2023). \doi{10.1109/CVPR52729.2023.01627}

\bibitem{discrete_shells}
Grinspun, E., Hirani, A.N., Desbrun, M., Schr\"{o}der, P.: Discrete shells. In: Proceedings of the 2003 ACM SIGGRAPH/Eurographics Symposium on Computer Animation. p. 62–67. SCA '03, Eurographics Association, Goslar, DEU (2003)

\bibitem{drape}
Guan, P., Reiss, L., Hirshberg, D.A., Weiss, A., Black, M.J.: Drape: Dressing any person. ACM Trans. Graph.  \textbf{31}(4) (jul 2012). \doi{10.1145/2185520.2185531}

\bibitem{garnet}
Gundogdu, E., Gundogdu, E., Constantin, V., Constantin, V., Seifoddini, A., Seifoddini, A., Dang, M., Dang, M., Dang, M.P., Salzmann, M., Salzmann, M., Fua, P., Fua, P.: Garnet: A two-stream network for fast and accurate 3d cloth draping. null  (2019). \doi{10.1109/iccv.2019.00883}

\bibitem{async_contact_mechanics}
Harmon, D., Vouga, E., Smith, B., Tamstorf, R., Grinspun, E.: Asynchronous contact mechanics. Commun. ACM  \textbf{55}(4),  102–109 (apr 2012). \doi{10.1145/2133806.2133828}

\bibitem{avatarCLIP}
Hong, F., Zhang, M., Pan, L., Cai, Z., Yang, L., Liu, Z.: Avatarclip: zero-shot text-driven generation and animation of 3d avatars. ACM Trans. Graph.  \textbf{41}(4) (jul 2022). \doi{10.1145/3528223.3530094}

\bibitem{simulating_knitted_cloth}
Kaldor, J.M., James, D.L., Marschner, S.: Simulating knitted cloth at the yarn level. In: ACM SIGGRAPH 2008 papers, pp.~1--9 (2008)

\bibitem{multi_res_pyramid}
Laczkó, H., Madadi, M., Escalera, S., Gonzalez, J.: A generative multi-resolution pyramid and normal-conditioning 3d cloth draping (2024)

\bibitem{estimating_cloth_params_with_position}
Larionov, E., Eckert, M.L., Wolff, K., Stuyck, T.: Estimating cloth elasticity parameters using position-based simulation of compliant constrained dynamics. arXiv.org  (2022). \doi{10.48550/arxiv.2212.08790}

\bibitem{texture_360}
Lazova, V., Lazova, V., Insafutdinov, E., Insafutdinov, E., Pons-Moll, G., Pons-Moll, G.: 360-degree textures of people in clothing from a single image. arXiv: Computer Vision and Pattern Recognition  (2019). \doi{10.1109/3dv.2019.00076}

\bibitem{clothcombo}
Lee, D., Kang, H., Lee, I.K.: Clothcombo: Modeling inter-cloth interaction for draping multi-layered clothes. ACM Trans. Graph.  \textbf{42}(6) (dec 2023). \doi{10.1145/3618376}

\bibitem{lee_and_lee}
Lee, D., Lee, I.K.: Multi-layered unseen garments draping network. arXiv preprint arXiv:2304.03492  (2023)

\bibitem{posespacedeform}
Lewis, J.P., Cordner, M., Fong, N.: Pose space deformation: A unified approach to shape interpolation and skeleton-driven deformation. In: Seminal Graphics Papers: Pushing the Boundaries, Volume 2. Association for Computing Machinery, New York, NY, USA (2023)

\bibitem{ipc}
Li, M., Ferguson, Z., Schneider, T., Langlois, T., Zorin, D., Panozzo, D., Jiang, C., Kaufman, D.M.: Incremental potential contact: intersection-and inversion-free, large-deformation dynamics. ACM Trans. Graph.  \textbf{39}(4) (aug 2020). \doi{10.1145/3386569.3392425}

\bibitem{cipc}
Li, M., Kaufman, D.M., Jiang, C.: Codimensional incremental potential contact. ACM Trans. Graph.  \textbf{40}(4) (jul 2021). \doi{10.1145/3450626.3459767}

\bibitem{dig}
Li, R., Guillard, B., Remelli, E., Fua, P.: Dig: Draping implicit garment over the human body. In: Wang, L., Gall, J., Chin, T.J., Sato, I., Chellappa, R. (eds.) Computer Vision -- ACCV 2022. pp. 344--359. Springer Nature Switzerland, Cham (2023)

\bibitem{isp}
Li, R., Guillard, B., Fua, P.: Isp: Multi-layered garment draping with implicit sewing patterns  (2023)

\bibitem{diffcloth}
Li, Y., Du, T., Wu, K., Xu, J., Matusik, W.: Diffcloth: Differentiable cloth simulation with dry frictional contact. ACM Trans. Graph.  \textbf{42}(1) (oct 2022). \doi{10.1145/3527660}

\bibitem{NCloth}
Li, Y., Li, Y., Tang, M., Tang, M., Yang, Y., Yang, Y., Huang, Z., Huang, Z., Tong, R., Tong, R., Yang, S., Yang, S., Li, Y., Li, Y., Manocha, D., Manocha, D.: N-cloth: Predicting 3d cloth deformation with mesh-based networks. null  (2021)

\bibitem{diffcloth_inverse_problems}
Liang, J., Lin, M., Koltun, V.: Differentiable cloth simulation for inverse problems. In: Wallach, H., Larochelle, H., Beygelzimer, A., d\textquotesingle Alch\'{e}-Buc, F., Fox, E., Garnett, R. (eds.) Advances in Neural Information Processing Systems. vol.~32. Curran Associates, Inc. (2019)

\bibitem{point_based_implicit}
Lin, S., Zhang, H., Zheng, Z., Shao, R., Liu, Y.: Learning implicit templates for point-based clothed human modeling. In: Avidan, S., Brostow, G., Ciss{\'e}, M., Farinella, G.M., Hassner, T. (eds.) Computer Vision -- ECCV 2022. pp. 210--228. Springer Nature Switzerland, Cham (2022)

\bibitem{stiffparams}
Liu, T., Bargteil, A.W., O'Brien, J.F., Kavan, L.: Fast simulation of mass-spring systems. ACM Trans. Graph.  \textbf{32}(6) (nov 2013). \doi{10.1145/2508363.2508406}

\bibitem{smpl}
Loper, M., Mahmood, N., Romero, J., Pons-Moll, G., Black, M.J.: {SMPL}: A skinned multi-person linear model. ACM Trans. Graphics (Proc. SIGGRAPH Asia)  \textbf{34}(6),  248:1--248:16 (Oct 2015)

\bibitem{drapenet}
Luigi, L.D., Li, R., Guillard, B., Salzmann, M., Fua, P.: Drapenet: Garment generation and self-supervised draping  (2023)

\bibitem{dlca}
Luo, C., Luo, F., Wang, Y., Zhao, E., Xiao, C.: Dlca-recon: Dynamic loose clothing avatar reconstruction from monocular videos (2023)

\bibitem{dry_frictional_contact}
Ly, M., Jouve, J., Boissieux, L., Bertails-Descoubes, F.: Projective dynamics with dry frictional contact. ACM Trans. Graph.  \textbf{39}(4) (aug 2020). \doi{10.1145/3386569.3392396}

\bibitem{learning_generative_clothing}
Ma, Q., Ma, Q., Ma, Q., Yang, J., Yang, J., Yang, J., Ranjan, A., Ranjan, A., Pujades, S., Pujades, S., Pons-Moll, G., Pons-Moll, G., Tang, S., Tang, S., Black, M.J., Black, M.J., Black, M.J.: Learning to dress 3d people in generative clothing. arXiv: Computer Vision and Pattern Recognition  (2019). \doi{10.1109/cvpr42600.2020.00650}

\bibitem{robust_sdf}
Macklin, M., Erleben, K., M\"{u}ller, M., Chentanez, N., Jeschke, S., Corse, Z.: Local optimization for robust signed distance field collision. Proc. ACM Comput. Graph. Interact. Tech.  \textbf{3}(1) (may 2020). \doi{10.1145/3384538}

\bibitem{amass}
Mahmood, N., Ghorbani, N., Troje, N.F., Pons-Moll, G., Black, M.J.: {AMASS}: Archive of motion capture as surface shapes. In: International Conference on Computer Vision. pp. 5442--5451 (Oct 2019)

\bibitem{occupancy_networks}
Mescheder, L., Mescheder, L., Oechsle, M., Oechsle, M., Niemeyer, M., Niemeyer, M., Nowozin, S., Nowozin, S., Geiger, A., Geiger, A.: Occupancy networks: Learning 3d reconstruction in function space. Computer Vision and Pattern Recognition  (2019). \doi{10.1109/cvpr.2019.00459}

\bibitem{data_driven_estimation}
Miguel, E., Bradley, D., Thomaszewski, B., Bickel, B., Matusik, W., Otaduy, M.A., Marschner, S.: Data-driven estimation of cloth simulation models. Comput. Graph. Forum  \textbf{31}(2pt2),  519–528 (may 2012)

\bibitem{nerf}
Mildenhall, B., Mildenhall, B., Srinivasan, P.P., Srinivasan, P.P., Tancik, M., Tancik, M., Barron, J.T., Barron, J.T., Ramamoorthi, R., Ramamoorthi, R., Ng, R., Ng, R.: Nerf: Representing scenes as neural radiance fields for view synthesis. European Conference on Computer Vision  (2020). \doi{10.1145/3503250}

\bibitem{adaptive_anis_remesh}
Narain, R., Samii, A., O'Brien, J.F.: Adaptive anisotropic remeshing for cloth simulation. ACM Trans. Graph.  \textbf{31}(6) (nov 2012). \doi{10.1145/2366145.2366171}

\bibitem{pointe}
Nichol, A., Jun, H., Dhariwal, P., Mishkin, P., Chen, M.: Point-e: A system for generating 3d point clouds from complex prompts. arXiv.org  (2022). \doi{10.48550/arxiv.2212.08751}

\bibitem{bone_motion_networks}
Pan, X., Mai, J., Jiang, X., Tang, D., Li, J., Shao, T., Zhou, K., Jin, X., Manocha, D.: Predicting loose-fitting garment deformations using bone-driven motion networks. In: ACM SIGGRAPH 2022 Conference Proceedings. SIGGRAPH '22, Association for Computing Machinery, New York, NY, USA (2022). \doi{10.1145/3528233.3530709}

\bibitem{deepsdf}
Park, J.J., Park, J.J., Florence, P., Florence, P.R., Straub, J., Straub, J., Newcombe, R., Newcombe, R., Lovegrove, S., Lovegrove, S.: Deepsdf: Learning continuous signed distance functions for shape representation. Computer Vision and Pattern Recognition  (2019). \doi{10.1109/cvpr.2019.00025}

\bibitem{tailornet}
Patel, C., Liao, Z., Pons-Moll, G.: Tailornet: Predicting clothing in 3d as a function of human pose, shape and garment style  (2020)

\bibitem{multiscale_graph_neural_networks}
Perera, R., Agrawal, V.: Multiscale graph neural networks with adaptive mesh refinement for accelerating mesh-based simulations. arXiv.org  (2024). \doi{10.48550/arxiv.2402.08863}

\bibitem{graphnet}
Pfaff, T., Fortunato, M., Sanchez-Gonzalez, A., Battaglia, P.W.: Learning mesh-based simulation with graph networks. ArXiv  \textbf{abs/2010.03409} (2020)

\bibitem{clothcap}
Pons-Moll, G., Pujades, S., Hu, S., Black, M.J.: Clothcap: seamless 4d clothing capture and retargeting. ACM Trans. Graph.  \textbf{36}(4) (jul 2017). \doi{10.1145/3072959.3073711}

\bibitem{mesh_tension}
Raman, C., Hewitt, C., Wood, E., Baltrušaitis, T.: Mesh-tension driven expression-based wrinkles for synthetic faces. IEEE Workshop/Winter Conference on Applications of Computer Vision  (2023). \doi{10.1109/wacv56688.2023.00351}

\bibitem{learning_to_measure}
Rasheed, A.H., Rasheed, A.H., Rasheed, A.H., Romero, V., Romero, V., Bertails-Descoubes, F., Bertails-Descoubes, F., Wuhrer, S., Wuhrer, S., Franco, J.S., Franco, J.S., Lazarus, A.J., Lazarus, A.: Learning to measure the static friction coefficient in cloth contact. null  (2020). \doi{null}

\bibitem{visual_approach}
Rasheed, A.H., Rasheed, A.H., Romero, V., Romero, V., Bertails-Descoubes, F., Bertails-Descoubes, F., Wuhrer, S., Wuhrer, S., Franco, J.S., Franco, J.S., Lazarus, A.J., Lazarus, A.: A visual approach to measure cloth-body and cloth-cloth friction. IEEE Transactions on Pattern Analysis and Machine Intelligence  (2021). \doi{10.1109/tpami.2021.3097547}

\bibitem{how_will_it_drape}
Rodríguez-Pardo, C., Prieto-Martin, M., Casas, D., Garcés, E.: How will it drape like? capturing fabric mechanics from depth images. Computer graphics forum (Print)  (2023). \doi{10.1111/cgf.14750}

\bibitem{complex_graphnets}
Sanchez-Gonzalez, A., Godwin, J., Pfaff, T., Ying, R., Leskovec, J., Battaglia, P.W.: Learning to simulate complex physics with graph networks. In: Proceedings of the 37th International Conference on Machine Learning. ICML'20, JMLR.org (2020)

\bibitem{unlef}
Santesteban, I., Otaduy, M., Thuerey, N., Casas, D.: Ulnef: Untangled layered neural fields for mix-and-match virtual try-on. In: Koyejo, S., Mohamed, S., Agarwal, A., Belgrave, D., Cho, K., Oh, A. (eds.) Advances in Neural Information Processing Systems. vol.~35, pp. 12110--12125. Curran Associates, Inc. (2022)

\bibitem{vto}
Santesteban, I., Otaduy, M.A., Casas, D.: Learning-based animation of clothing for virtual try-on  (2019)

\bibitem{snug}
Santesteban, I., Otaduy, M.A., Casas, D.: Snug: Self-supervised neural dynamic garments  (2022)

\bibitem{self_supervised_collision_vto}
Santesteban, I., Thuerey, N., Otaduy, M.A., Casas, D.: Self-supervised collision handling via generative 3d garment models for virtual try-on. In: 2021 IEEE/CVF Conference on Computer Vision and Pattern Recognition (CVPR). pp. 11758--11768 (2021). \doi{10.1109/CVPR46437.2021.01159}

\bibitem{garment3dgen}
Sarafianos, N., Stuyck, T., Xiang, X., Li, Y., Popovic, J., Ranjan, R.: Garment3dgen: 3d garment stylization and texture generation. null  (2024). \doi{null}

\bibitem{usr}
Shi, Y., Xiong, Y., Chai, J., Ni, B., Zhang, W.: Usr: Unsupervised separated 3d garment and human reconstruction via geometry and semantic consistency  (2023)

\bibitem{drape_sim}
Shim, E., Ju, E., Choi, M.G.: Drape simulation estimation for non-linear stiffness model. null  (2023). \doi{10.15701/kcgs.2023.29.3.117}

\bibitem{deepCloth}
Su, Z., Yu, T., Wang, Y., Liu, Y.: Deepcloth: Neural garment representation for shape and style editing. IEEE Transactions on Pattern Analysis and Machine Intelligence  \textbf{45}(2),  1581--1593 (2023). \doi{10.1109/TPAMI.2022.3168569}

\bibitem{gnn_physics_engine}
Álvaro Sánchez‐González, Sanchez-Gonzalez, A., Sanchez, A., Heess, N., Heess, N., Springenberg, J.T., Springenberg, J.T., Merel, J., Merel, J., Riedmiller, M., Hadsell, R., Hadsell, R., Riedmiller, M., Battaglia, P.W., Battaglia, P.W.: Graph networks as learnable physics engines for inference and control. null  (2018). \doi{null}

\bibitem{icloth}
Tang, M., wang, t., Liu, Z., Tong, R., Manocha, D.: I-cloth: incremental collision handling for gpu-based interactive cloth simulation. ACM Trans. Graph.  \textbf{37}(6) (dec 2018). \doi{10.1145/3272127.3275005}

\bibitem{asynchronous_cloth}
Thomaszewski, B., Pabst, S., Stra{\ss}er, W.: Asynchronous cloth simulation (2008)

\bibitem{neural_gif}
Tiwari, G., Sarafianos, N., Tung, T., Pons-Moll, G.: Neural-gif: Neural generalized implicit functions for animating people in clothing. IEEE International Conference on Computer Vision  (2021). \doi{10.1109/iccv48922.2021.01150}

\bibitem{garsim}
Tiwari, L., Bhowmick, B.: Garsim: Particle based neural garment simulator. null  (2023). \doi{10.1109/wacv56688.2023.00445}

\bibitem{gensim}
Tiwari, L., Bhowmick, B., Sinha, S.: Gensim: Unsupervised generic garment simulator. null  (2023). \doi{10.1109/cvprw59228.2023.00439}

\bibitem{real_time_radiance_field}
Trevithick, A., Chan, M., Stengel, M., Chan, E., Liu, C., Yu, Z., Khamis, S., Chandraker, M., Ramamoorthi, R., Nagano, K.: Real-time radiance fields for single-image portrait view synthesis. ACM Trans. Graph.  \textbf{42}(4) (jul 2023). \doi{10.1145/3592460}

\bibitem{graph_3D_point}
Valsesia, D., Valsesia, D., Fracastoro, G., Fracastoro, G., Magli, E., Magli, E.: Learning localized generative models for 3d point clouds via graph convolution. International Conference on Learning Representations  (2019). \doi{null}

\bibitem{non_linear_stiffness}
Volino, P., Magnenat-Thalmann, N., Faure, F.: A simple approach to nonlinear tensile stiffness for accurate cloth simulation. ACM Trans. Graph.  \textbf{28}(4) (sep 2009). \doi{10.1145/1559755.1559762}

\bibitem{gpu_based}
Wang, H.: Gpu-based simulation of cloth wrinkles at submillimeter levels. ACM Trans. Graph.  \textbf{40}(4) (jul 2021). \doi{10.1145/3450626.3459787}

\bibitem{data_driven_elastic_model}
Wang, H., O'Brien, J.F., Ramamoorthi, R.: Data-driven elastic models for cloth: modeling and measurement. In: ACM SIGGRAPH 2011 Papers. SIGGRAPH '11, Association for Computing Machinery, New York, NY, USA (2011). \doi{10.1145/1964921.1964966}

\bibitem{meta_avatar}
Wang, S., Mihajlovic, M., Ma, Q., Geiger, A., Tang, S.: Metaavatar: Learning animatable clothed human models from few depth images. Neural Information Processing Systems  (2021). \doi{null}

\bibitem{cloning_outfits}
Wang, Y., Liang, X., Liao, S.: Cloning outfits from real-world images to 3d characters for generalizable person re-identification  (2022)

\bibitem{graph_based_synthesis}
Weiß, S., Moulin, J.P., Chandran, P., Zoss, G., Gotardo, P., Bradley, D.: Graph‐based synthesis for skin micro wrinkles. Computer graphics forum (Print)  (2023). \doi{10.1111/cgf.14904}

\bibitem{icon}
Xiu, Y., Yang, J., Tzionas, D., Black, M.J.: Icon: Implicit clothed humans obtained from normals. Computer Vision and Pattern Recognition  (2021). \doi{10.1109/cvpr52688.2022.01294}

\bibitem{learning_based_recovery}
Yang, S., Yang, S., Liang, J., Liang, J., Lin, M.C., Lin, M.C.: Learning-based cloth material recovery from video. null  (2017). \doi{10.1109/iccv.2017.470}

\bibitem{point_based_modeling}
Zakharkin, I., Mazur, K., Grigorev, A., Lempitsky, V.: Point-based modeling of human clothing  (2021)

\bibitem{lion}
Zeng, X., Zeng, X., Vahdat, A., Vahdat, A., Williams, F., Williams, F.H., Gojcic, Z., Žan Gojčič, Litany, O., Litany, O., Fidler, S., Fidler, S., Kreis, K., Kreis, K.: Lion: Latent point diffusion models for 3d shape generation. Cornell University - arXiv  (2022). \doi{10.48550/arxiv.2210.06978}

\bibitem{avatargen}
Zhang, J., Jiang, Z., Yang, D., Xu, H., Shi, Y., Song, G., Xu, Z., Wang, X., Feng, J.: Avatargen: a 3d generative model for animatable human avatars  (2022)

\bibitem{diff_material_robotic_fish}
Zhang, J.Z., Zhang, Y., Ma, P., Nava, E., Du, T., Arm, P., Matusik, W., Katzschmann, R.K.: Learning material parameters and hydrodynamics of soft robotic fish via differentiable simulation. arXiv.org  (2021). \doi{null}

\bibitem{motion_guided}
Zhang, M., Ceylan, D., Mitra, N.J.: Motion guided deep dynamic 3d garments. ACM Trans. Graph.  \textbf{41}(6) (nov 2022). \doi{10.1145/3550454.3555485}

\end{thebibliography}
\end{document}